\documentclass[10pt]{article}
\usepackage[utf8]{inputenc}
\usepackage[margin=1.0in]{geometry}
\usepackage{epsfig}
\usepackage{float}
\usepackage{graphics}
\usepackage{subfig}
\usepackage{comment}
\usepackage{colortbl}
\usepackage{colordvi}
\usepackage{breqn}
\usepackage{multicol}
\usepackage{amsmath}
\usepackage{cite}
\title{Self-organized intracellular twisters}
\author{Sayantan Dutta$^{1,2,\#}$, Reza Farhadifar$^{2,\#}$, Wen Lu$^{3}$, Gokberk Kabacaoğlu$^{2}$, Robert Blackwell$^{2}$ \\David B. Stein$^2$, Margot Lakonishok$^{3}$, Vladimir I. Gelfand$^{3}$, Stanislav Y. Shvartsman$^{1,2,4}$ \\ Michael J. Shelley$^{2,5}$}
\date{%
    $^1$Department of Chemical and Biological Engineering, Princeton University, Princeton, NJ\\%
    $^2$Center of Computational Biology, Flatiron Institute, New York, NY\\
    $^3$Department of Cell and Developmental Biology, Feinberg School of Medicine, Northwestern University, Chicago, IL\\
    $^4$Department of Molecular Biology and Lewis Sigler Institute of Integrative Genomics, Princeton University, Princeton, NJ\\
    $^5$Courant Institute of Mathematical Sciences, New York University, New York, NY\\
    $^\#$equal contribution
    }
\begin{document}
\bibliographystyle{naturemag}
\maketitle
  
{\it Life in complex systems, such as cities and organisms, comes to a standstill when global coordination of mass, energy, and information flows is disrupted. Global coordination is no less important in single cells, especially in large oocytes and newly formed embryos, which commonly use fast fluid flows for dynamic reorganization of their cytoplasm. Here, we combine theory, computing, and imaging to investigate such flows in the \textit{Drosophila} oocyte, where streaming has been proposed to spontaneously arise from hydrodynamic interactions among cortically anchored microtubules loaded with cargo-carrying molecular motors. We use a fast, accurate, and scalable numerical approach to investigate fluid-structure interactions of 1000s of flexible fibers and demonstrate the robust emergence and evolution of cell-spanning vortices, or twisters. Dominated by a rigid body rotation and secondary toroidal components, these flows are likely involved in rapid mixing and transport of ooplasmic components.}

\smallskip
Cytoplasmic streaming, first described in 1774 by Bonaventura Corti~\cite{Corti1774}, regulates a wide range of intracellular processes, especially when large cell size makes diffusion and motor-driven transport too slow for efficient intracellular transport and mixing ~\cite{Yi2011, Almonacid2015, Deneke2019, Glotzer1997, Van2008, Hird1993, Emmons1995, Trong2015, Gross2019}. The types of cytoplasmic flows can vary significantly among cells and over different stages of cell development, from random streaming in early fly oocytes to circulation flows observed in plants and later oocytes, to shuttle streaming found in slime molds~\cite{Goldstein2015,LG2023}. Cytoplasmic flows are commonly driven by forces originating from the cell cortex, where motor proteins carry cargo along cytoskeletal filaments and so entrain the fluid~\cite{shamipour2021}. The formation of macroscopic flows in cells requires an alignment in motor movements across many filaments. In some cells, such as the algae {\it Chara}, this alignment originates from the organization of actin filaments at earlier developmental stages, and can be viewed as providing static boundary conditions for the flow problem~\cite{Van2008,Woodhouse2013}. In others, like oocytes of the fruit fly \textit{Drosophila}, our focus here, motors move upon a cytoskeletal bed of flexible microtubules, and their alignment and direction was proposed to arise via self-amplifying feedback between motor-induced cytoplasmic flows and collective deformations of the microtubule bed~\cite{Quinlan2016}. Understanding how these large-scale flows emerge in a system of hydrodynamically coupled deformable fibers is highly nontrivial due to strong geometric nonlinearities and widely separated spatiotemporal scales. Here, we present a versatile modeling approach for tackling this challenge and show how it can be combined with experiments in the \textit{Drosophila} oocyte to provide general insights into self-organized cytoplasmic streaming.   

The cytoplasm of the developing \textit{Drosophila} oocyte remains relatively quiescent for the first three days of oogenesis. During this time, diffusion and directed transport are used to localize several molecular factors needed for the patterning of the future embryo~\cite{becalska2009,Trong2015}. Later, when the oocyte is 150-300 $\mu m$ long and 100-200 $\mu m$ wide, large-scale streaming arises, often appearing as a vortex, and having a typical speed of 100-400 nm/s~\cite{Gutzeit1982,Monteith2016,Lu2016}. This vortex was proposed to be generated by beds of cortically anchored flexible microtubules serving as tracks for plus-end-directed Kinesin-1 motor proteins moving free microtubules and other payloads~\cite{Palacios2002, Serbus2005, Ganguly2012, Monteith2016, Lu2016,Ravichandran2019}. Recently, a coarse-grained model, based on an active and deformable porous-medium model, was used to argue that coherent vortical flows, hundreds of microns in size, can self-organize via fluid-mediated coupling of active and flexible microtubules ~\cite{Stein2021}. Parametric analysis of special, azimuthally homogeneous solutions in a 2D disk geometry, identified a regime where all microtubules coherently bend due to motor activity, resulting in a large-scale vortical flow. While the model supports the idea of a self-amplifying feedback generating large-scale flows, it remains to be determined whether and how this mechanism works in more realistic 3D geometries where homogeneous solutions are not allowed, and when other simplifying features of the model are removed. We present a computational approach which allows us to address these questions while making new and testable experimental predictions about the 3D structure of cytoplasmic flows. 

\begin{figure} [tbh!]
\centering
\includegraphics[width=16cm]{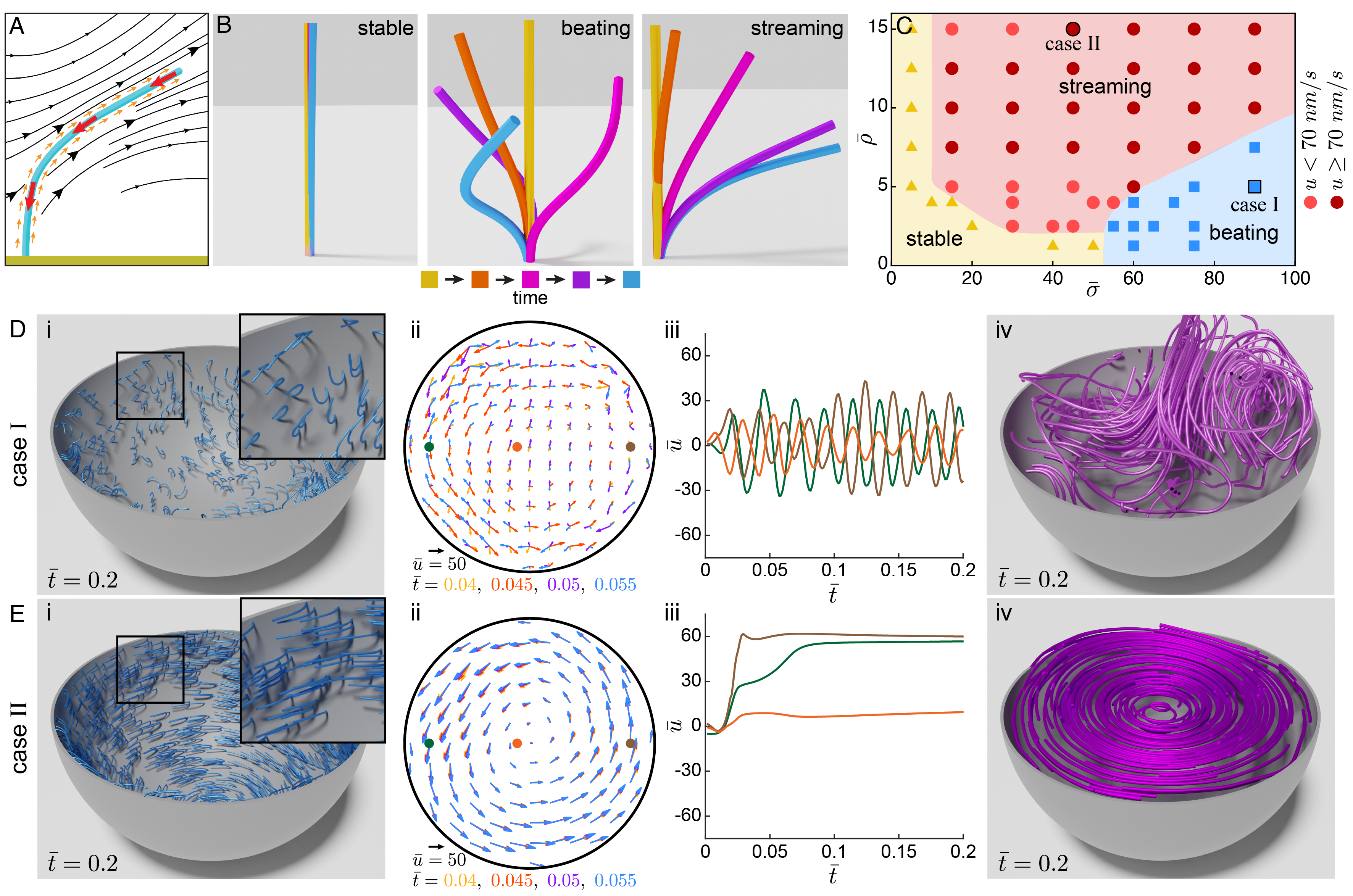}
\caption{\textbf{Hydrodynamic interactions of motor-loaded microtubules generate intracellular flows.} (A)~A schematic illustrating a microtubule (blue) anchored normal to the cell surface (green) subject to active forcing and immersed in a viscous Newtonian fluid (i.e. cytoplasm). Large red arrows show compressive stress on the microtubule, small orange arrows represent stress on the fluid, and black lines indicate flows in the fluid. (B)~Three regimes of microtubule behavior: stable regime with little microtubule deformation (left), beating regime with microtubules oscillating (middle; case I), and the streaming regime with microtubules (collectively) bending (right; case II). Microtubule colors represent time evolution. (C)~A phase diagram of microtubule behaviors as a function of adimensional microtubule areal density $\bar{\rho}$ and motor forces $\bar{\sigma}$. The regions in yellow, red, and blue represent stable, streaming, and oscillation phases, respectively. The color of the red circles represents the characteristic streaming speed (flow speed at a distance $0.8R$ from the center). The Case I data point has $\bar{\rho}$=5 and $\bar{\sigma}$=90, while case II has $\bar{\rho}=15$ and $\bar{\sigma}=45$. (D,E)~For cases I and II, cut-away view of instantaneous microtubule configurations in the spherical cell; (i) 2D projection of velocity field in the sectioning equatorial plane, at four time points; (ii) the adimensional azimuthal velocity component at the three points in the equatorial plane, as a function of $\bar t$; (iii) and 3D streamlines integrated from the 3D velocity field (iv). Curve colors correspond to the colored point in preceding velocity plots.}
\label{FIG1}
\end{figure}

\smallskip
{\it Modeling hydrodynamically coupled motor-driven fibers:} 
Conceptually, plus-end directed motors -- here, kinesin-1 -- bind along bound microtubules, carrying cargos towards free plus-ends and detaching once they reach there. While the cargos appear various, possibly including cellular organelles and yolk granules, free microtubules have been identified as one cargo crucial for robust streaming \cite{Lu2016}. Given the lack of data on the details of cargo binding, and payload densities and sizes, we assume the simplest model and, as in \cite{Stein2021}, coarse-grain the forces of plus-end directed cargos upon bound microtubules to a uniform compressive force density (i.e. directed along the bound microtubule toward its anchored end) (Fig.~\ref{FIG1}A). An equal and opposite force is exerted upon the surrounding fluid, thus satisfying Newton's Third Law. We find this is sufficient to recover many aspects of observed streaming, and to make several predictions for experiment.

Hence, consider $N$ microtubules clamped to the inner surface of a spheroidal cell of effective radius $R$ and surface area $S~(\sim R^2)$. Microtubules are well modeled as inextensible elastic slender bodies ($\epsilon=$~microtubule radius/length $\sim10^{-3}\ll 1$)~\cite{gittes1993, shelley2016}, and the cytoplasm is modeled as a Newtonian fluid of viscosity $\mu$ \cite{Ganguly2012}. The shape of microtubule $i$ at time $t$ is given by $\mathbf{X}^i(s,t)$, where $0\leq s \leq L^i$ is the arclength from its base and $L^i$ its length. Microtubule shape evolution, due to drag forces balancing elastic and motor forces, is given by local slender-body theory~
\cite{KR1976,Tornberg2004, Nazockdast2017}
\begin{equation}
     \label{MT_model1}
    \eta({\mathbf{X}}_t^i-\bar{\mathbf{u}}^i(\mathbf{X}^i))=(\mathbf{I}+\mathbf{X}^i_s\mathbf{X}^i_s)(\mathbf{f}^i-\sigma\mathbf{X}^i_s),
\end{equation}
where $\mathbf{X}_s^i=\partial_s\mathbf{X}^i$ is the microtubule unit tangent vector, and $\eta=8\pi \mu/c(\epsilon)$ a drag coefficient, having $c=|\ln e \epsilon^2|$ (given the logarithm, changes in $\epsilon$ enter very weakly). The velocity $\bar{\mathbf{u}}^i$ is that induced by all other microtubules and backflow from the cortex. The force density $\mathbf{f}^i=-E\mathbf{X}^i_{ssss}+(T^i\mathbf{X}^i_s)_s$ is the elastic force due to microtubule bending, with rigidity $E$, and tensile forces, with tension $T^i$ enforcing inextensibility. The microtubule aligned term $-\sigma\mathbf{X}^i_s$ is the coarse-grained compressive load ($\sigma>0$) exerted by kinesin-cargo complexes. 

Given the background fluid velocity, Eq.~(1) describes an initial value problem for microtubule shape. The background cytoplasmic velocity $\mathbf{u}(\mathbf{x})$, induced by all microtubules and by periphery backflow, satisfies the forced Stokes equation:
\begin{equation}
    \label{MT_model2}
   \nabla p-\mu\Delta\mathbf{u}=\sum_{i=1}^{N}\int_0^{L^i}ds~\mathbf{f}^i(s)\delta(\mathbf{x}-\mathbf{X}^i(s)); ~\mathbf{\nabla}\cdot\mathbf{u}=0,
\end{equation}
where $p$ is the pressure and no-slip is taken on the cortex. The minus-ends of microtubules are pinned and clamped (respectively, $\mathbf{X}(0,t)=\mathbf{X}(0,0)$ and $\mathbf{X}_s(0,t)=-\hat{\mathbf{n}}(\mathbf{X}(0,t))$ with $\hat{\mathbf{n}}$ the outward surface normal) at the cortex, and the free plus-end taken as torque and force free. That motor forces do not show up directly in determining the background velocity reflects their subdominant dipolar nature and the assumed close proximity of payloads to the load-bearing microtubules.

Equations (1) \& (2) reflect a multiscale structure, with Eq.~(1) evolving individual microtubules moving in a background flow created, via Eq.~(2) by the collective forcing of the microtubule ensemble. Beds of motile cilia, also a multiscale active polymer transport system, are much studied for their capacity to self-organize, including through hydrodynamic interactions, into metachronal waves \cite{CFS2022}. Quite unlike the system studied here, cilia are internally actuated by dynein motors moving upon ciliary microtubule doublets with fluid motion created directly by ciliary motion. Not so here where even a single stationary and straight clamped microtubule will produce a upward cytoplasmic flow around it as a consequence of the payloads moving up on it, as conceptualized by Eqs.~(1) \& (2).

\smallskip
{\it Control parameters and numerical approach:} The parameters in the model combine to yield three important time-scales. Letting $L$ be a characteristic microtubule length, from Eq. (1) arises $\tau_r=\eta L^4/E$, the relaxation time of a single microtubule, and $\tau_m=L\eta/\sigma$, the time for motor forces to move a microtubule its own length. Equation (2), yields a second, faster microtubule relaxation time arising from collective hydrodynamic interactions, $\tau_c=\tau_r/\bar{\rho}$ where $\bar{\rho}=8\pi NL^2/cS$ is the effective areal density of microtubules \cite{Stein2021}. Ratios of these time-scales determine the two dimensionless parameters dependent upon on the biophysical properties of microtubules and motors and their numbers: the dimensionless microtubule areal density  $\bar{\rho}=\tau_r/\tau_c$, already introduced, and the dimensionless motor force $\bar{\sigma}=\tau_r/\tau_m=\sigma L^3/E$. The model has only two other, geometric, parameters: the ratio of microtubule length to system size, $\delta=L/R$, and $\epsilon$ (entering weakly). Here we keep $\delta$ and $\epsilon$ constant, and thus, $\bar{\rho}$ and $\bar{\sigma}$ govern the behavior of the system. 

Simulating this system efficiently for thousands of microtubules has its peculiar challenges. The microtubules make the system geometry very complex while their shape evolution is stiff due to their elasticity. While the number of degrees of freedom -- mainly discretized microtubule forces and shapes -- is not extreme ($\sim 10^{5-6}$), all are globally coupled by the Stokes equations, and the system need to be simulated to long times. For this, we developed an fast and scalable computational platform that accurately evolves Eqs. (1) and (2). It has three major components: First, boundary integral representations and slender body theory reduce the 3D Stokes equations in this complex domain to solving 1D integro-differential equations on microtubules and a coupled 2D integral equation on cell surfaces of nearly arbitrary geometry \cite{Nazockdast2017}. Second, a fast Stokes solver efficiently evaluates the nonlocal hydrodynamic interactions between microtubules and the periphery with linear scaling in the number of unknowns. Third, we use a stable implicit-explicit time-stepping scheme to efficiently evolve the stiff microtubules dynamics. This open source software is modular allowing parallel computations across multiple nodes \cite{SkellySim}. 

\smallskip
{\it Self-organized regimes in a spherical cell:} This infrastructure allows us to determine whether and how the behaviors predicted by the 2D analysis of the active porous medium model survive in a fully 3D geometry which, for example, disallows homogeneous solutions. Abstracting the cell to a sphere, we studied the model's long-time behavior for various combinations of $\bar{\rho}$ and $\bar{\sigma}$. For a given $\bar{\rho}$, we placed the microtubules at random, statistically uniform positions with initial straight configurations normal to the surface. We found three basic microtubule behaviors: microtubules that stay nearly straight, microtubules that beat with near periodicity, and microtubules that bend and remain bent~(Fig.~\ref{FIG1}B). We grouped the simulations based on these behaviors and used this classification to map out the domains of qualitatively different behaviors; see~(Fig.~\ref{FIG1}C). We find a stable phase for low motor forces or low density where microtubules remain nearly straight (Fig.~\ref{FIG1}C, yellow). 
For larger motor forces and moderate microtubule density, we observed phases where microtubules periodically beat (~Fig.~\ref{FIG1}C, blue). For a large range of parameters where microtubule density and motor strength are balanced, the model exhibits the streaming phase, where most microtubules bend collectively and remain bent (Fig.~\ref{FIG1}C, red). Consistent with the prediction of the earlier coarse-grained analysis, the beating regime arises only beyond a critical motor force, and streaming regimes arise only above a critical microtubule density.

The microtubules' configurations and forces determine the instantaneous 3D flow structure through Eq.~(2). In the stable regime, there are local cytoplasmic flows near each microtubule but negligible flows inside the cell~(see supplementary Fig.~S1). For case I in the phase diagram, initially straight microtubules evolve into beating states and oscillate with short-range and time-varying phase synchrony~[Fig.~\ref{FIG1}D(i) inset and supplementary Movie 1]. The associated internal velocity field is spatially complex and unsteady [Fig.~\ref{FIG1}D(ii) and supplementary Movie 2]. Individual velocity traces show a basic underlying frequency [e.g. Fig.~\ref{FIG1}D(iii)] on the scale of $\tau_m$, the motor time, at essentially the beat frequency of isolated active microtubule (not shown), but with persistent relative phase drift. The spatial complexity of these flows is revealed in Fig.~\ref{FIG1}d(iv) which shows the instantaneous streamlines within the cell. Perhaps coincidentally, the short range spatial correlation is reminiscent of cytoplasmic {\it seething} in earlier development stages\cite{Theurkauf1994, Serbus2005, Dahlgaard2007}. For case II in the phase diagram, we find that the initially straight microtubules at first bend in seemingly random directions. Then, gradually, these deformations align together into an array of bent microtubules wrapping around an axis of symmetry [Fig.~\ref{FIG1}E(i), and supplementary Movie 3]. This emergent axis is sensitive to details of initial data and microtubule patterning. The associated streaming flow is nearly steady, strongly vortical, spans the cell, has speeds $\sim$ 100nm/s, and is reached rapidly on the order of $\tau_c$, the collective relaxation time [Figs.~\ref{FIG1}E(ii,iii,iv), and supplementary Movie 4]. 

\begin{figure}[t!]
\centering
\includegraphics[width=8.5cm]{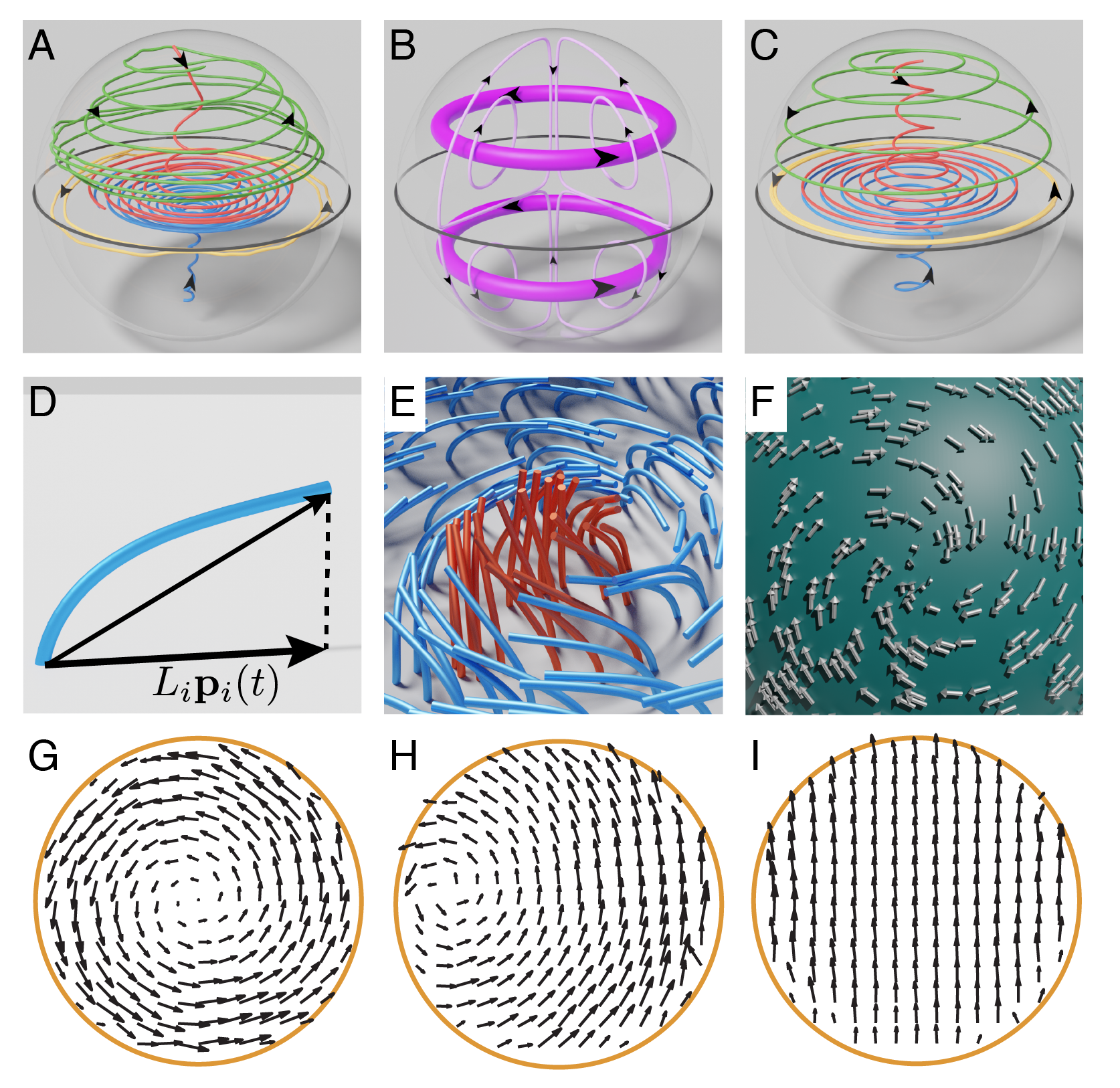}
\caption{\textbf{Twisters are a combination of a strong vortical flow and a weak bitoroidal flow.} (A)~Streamlines from a simulation of case II in the streaming region. Streamlines starting near the equatorial plane remain there, showing nearly circular paths (yellow). Streamlines starting near the poles move inwards on a spiral path towards the equatorial plane and then expand toward the periphery (red and blue). Streamlines starting above (or below) the equatorial plane near the cell periphery show the return flow back towards the pole (green). (B) Streamlines from two simple solutions to the Stokes equations inside a sphere: a strong 2D constant vorticity flow (thick streamlines) and a weaker bitoroidal flow (thin streamlines). Both velocity fields are tangent to the confining sphere. (C) Streamlines from best-fit combination of these two flows in approximating the flow in (A). (D)~A schematic defining the microtubule polarity vector $\mathbf p$. The blue fiber represents the configuration of a microtubule anchored to the surface. The black vectors represent the end-to-end vector in 3D and its projection $\mathbf{p}$ on the surface. Note $|\mathbf{p}|\leq 1$. (E) Microtubules configuration near the defect for simulation in (A). Microtubules with $p<0.7$ (less bent) are colored in red, and those with $p\geq 0.7$ are colored blue. (F)~ View of the microtubule polarity vector-field $p$ at the defect center shown in (E) with low microtubule alignment. (G,H,I) 2D projection of velocity fields from simulation in (A) in the sectioning equatorial planes when the plane normal is aligned with the z-axis (G), has angle $\pi/4$ relative to the z-axis (H), and is perpendicular to the z-axis (I).}
\label{FIG2}
\end{figure}

Setting the axis of the twister flow to $\hat{\mathbf{z}}$ and examining its 3D streamlines reveals its vortical component as nearly filling the cell, and having a weaker swirl component (Figs.~\ref{FIG2}A,B) moving fluid inwards from the poles along the twister axis with a return flow outwards along the cell perimeter. That is, the secondary flow consists of two counter-rotating toroidal vortices, one in each hemisphere, wrapped around the vortical axis. The simulated streaming flows can be well-fit away from the boundaries as a flow with surface slip by superposing a purely 2D rotational flow with an aligned axisymmetric bitoroidal flow that satisfies the no-slip condition\cite{stone1991}. For case II the strength of the rotational flow $\sim$20 times larger than the bitoroidal~(Figs.~\ref{FIG2}B,C, and supplementary information).

The dominant vortical flow arises from the collective bending of the majority of microtubules around a common axis [Fig.~\ref{FIG1}E(i)], while the swirling flow arises from microtubule conformations near "defect centers", our label for the points where the internal vortex axis ends at the cell surface. As a proxy for microtubule orientation and deformation we define the microtubule surface polarity vector (Fig.~\ref{FIG2}D):
\begin{equation}
    \mathbf p_i(t)=(\mathbf{I}-\hat{\mathbf{n}}\hat{\mathbf{n}})\cdot\frac{\mathbf{X}^i(L^i,t)-\mathbf{X}^i(0,t)}{L^i},~~\mbox{giving that}~~|\mathbf{p}_i|\leq 1.
\end{equation} 
Microtubules near defect centers are relatively straight (i.e. have low $|p_i|$), as the vortical flow direction becomes indeterminant there (Fig.~\ref{FIG2}E). Consequently, payloads moving upwards upon these microtubules produce an inwards force, creating the axial flow which underlies the secondary flow. Incompressibility yields the return flow. In defect regions (Fig.~\ref{FIG2}F) the $\mathbf{p}$ field  shows an inward spiraling patterning, consistent with the rotational flow and the defect pumping of cytoplasm which pulls fluid in peripherally, bending microtubules towards the defect center. In the language of liquid crystal physics, the polarity field structure is a combination of two $+$1-order disclination  singularities.

The twisters we find here are cell-spanning, three dimensional flow states. This has consequences for their observation via microscopy, which typically images 2D cross-sections slicing through the cell (though confocal z-stacks may give some 3D structure). One then expects that the flows thus imaged will be in planar cross-sections set at some random angle relative to the twister axis. This is illustrated in Figs.~\ref{FIG2}G-I which, by sampling in differently angled planes through the cell, show a full vortex, an apparently displaced and distorted one, and a fully transverse streaming flow. These are nonetheless all images of the same flow state.

\begin{figure}[b!]
\centering
\includegraphics[width=15cm]{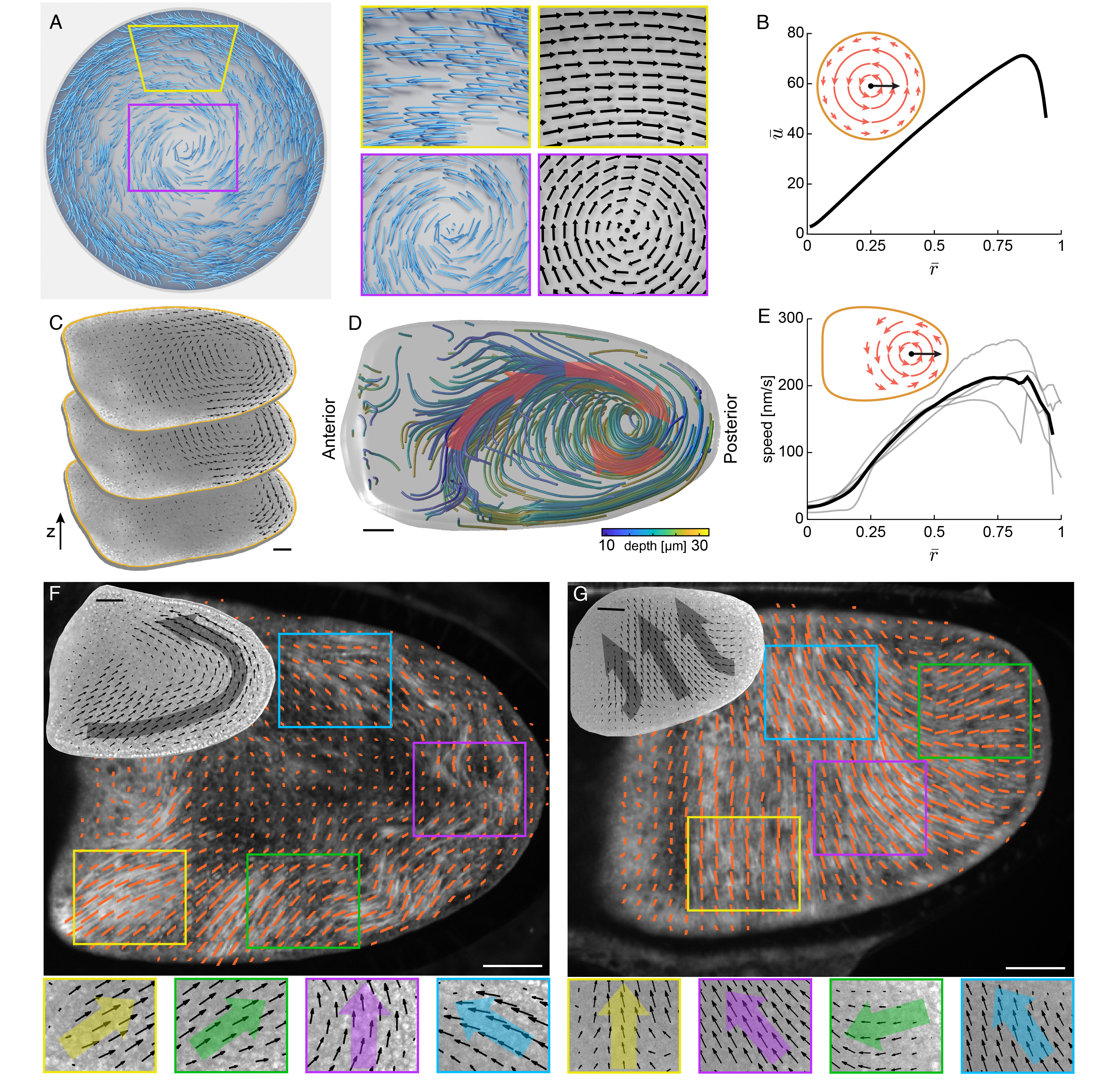}
\caption{\textbf{Surface microtubule orientation and cytoplasmic velocity fields.} (A)~Microtubule configurations at steady state from a simulation on a sphere from the streaming region (left) and selected regions from the same snapshot with corresponding velocity fields in the vicinity of the surface (right). (B)~Adimensional speed as a function of normalized distance $\bar{r}=r/R$ from the vortex center in a cross-section (shown as inset) for simulation in (A). (C)~2D velocity field in an oocyte measured by overset grid particle image velocimetry. Arrows show the direction and relative magnitude of the velocity. Measurements were done at consecutive z-sections from the oocyte surface. Scale bar, $25\mu m$. (D)~Computed 3D streamlines near the oocyte surface from the reconstructed velocity field. The color indicates depth. Red arrows show the flow direction. Scale bar, $25\mu m$.  (E)~Flow speed as a function of normalized distance $\bar{r}$ from the vortex center (shown as inset) measured in oocytes. Here, $\bar{r}=r/r_0$, where $r_0$ is the shortest distance from the vortex center to the periphery. Gray lines indicate measurements from four different oocytes, and the black line shows their average. (F,G)~Near-surface microtubules in oocytes imaged in maternally derived GFP-$\alpha$tub \textit{Drosophila}. Orange lines represent microtubules' local orientation with length representing the local degree of microtubules alignment measured by Gabor filter response. The cytoplasmic velocity field (black arrows) measured by PIV in the same oocyte in a plane $15\mu m$ from the surface for corresponding regions are shown. The colored arrows are average microtubules orientation in the corresponding regions. Scale bars, $25\mu m$.}
\label{FIG3}
\end{figure}

\smallskip
{\it Live imaging of cytoplasmic flows and cortical microtubules:} A central feature of our model is that the flow near the cell surface is locally set by the orientation of the microtubules ~(Fig.~\ref{FIG3}A). Moreover, because the flow is generated by motors moving along the microtubules, the flow speed increases from the center of the vortex towards the microtubule bed, then diminishes near the cell surface due to the no-slip boundary condition~(Fig.~\ref{FIG3}B). We successfully tested both of these predictions in live \textit{Drosophila} oocytes (supplementary Movie 5). We performed particle image velocimetry (PIV) using endogenous particles (likely yolk granules and other particles) as flow tracers. To accurately measure the 3D cytoplasmic flow field both in the interior and near the oocyte cortex this we used overset grids, which combine square grids in the interior of the oocytes with surface-conforming grids near the cortex derived from its local geometry~(Fig.~\ref{FIG3}C, supplementary information and Fig. S2). We often observed a vortical flow spanning $\sim$100 $\mu m$, with characteristic flow speeds of 100-300 nm/s~(Fig.~\ref{FIG3}D, supplementary Fig. S3). Our flow measurements are similar to previous studies~\cite{Quinlan2016, Williams2014} and comparable with the walking speed of Kinesin-1 ($\sim$200-500 nm/s)~\cite{Brendza1999,loiseau2010,Monteith2016,Lu2016}.
Consistent with modeling predictions, the speed of cytoplasmic flow increased from the center of the vortex towards the oocyte periphery and sharply decreased near the cortex~(Fig.~\ref{FIG3}E). The measured flow speed (100-300 nm/s) is comparable to simulations ($\sim$100 nm/s). We measured the local microtubule orientation in confocal fluorescent images of oocytes expressing GFP-tagged $\alpha$-tubulin (Fig. 3F, G, supplementary Fig. S4, supplementary information). In agreement with the model, the local cytoplasmic velocity field in oocytes is well-aligned with the orientation field of cortical microtubules; see Figs.~\ref{FIG3}F, G. 

These two flow reconstructions, which show a distorted vortical structure and transverse streaming, and Fig.~\ref{FIG3}D which shows a clear vortex, illustrate the variety of flows actually observed through live imaging. They are consistent with Figs.~\ref{FIG2}G,H,I as image slices through a basic twister structure.

\begin{figure}[h!]
\centering
\includegraphics[width=14.5cm]{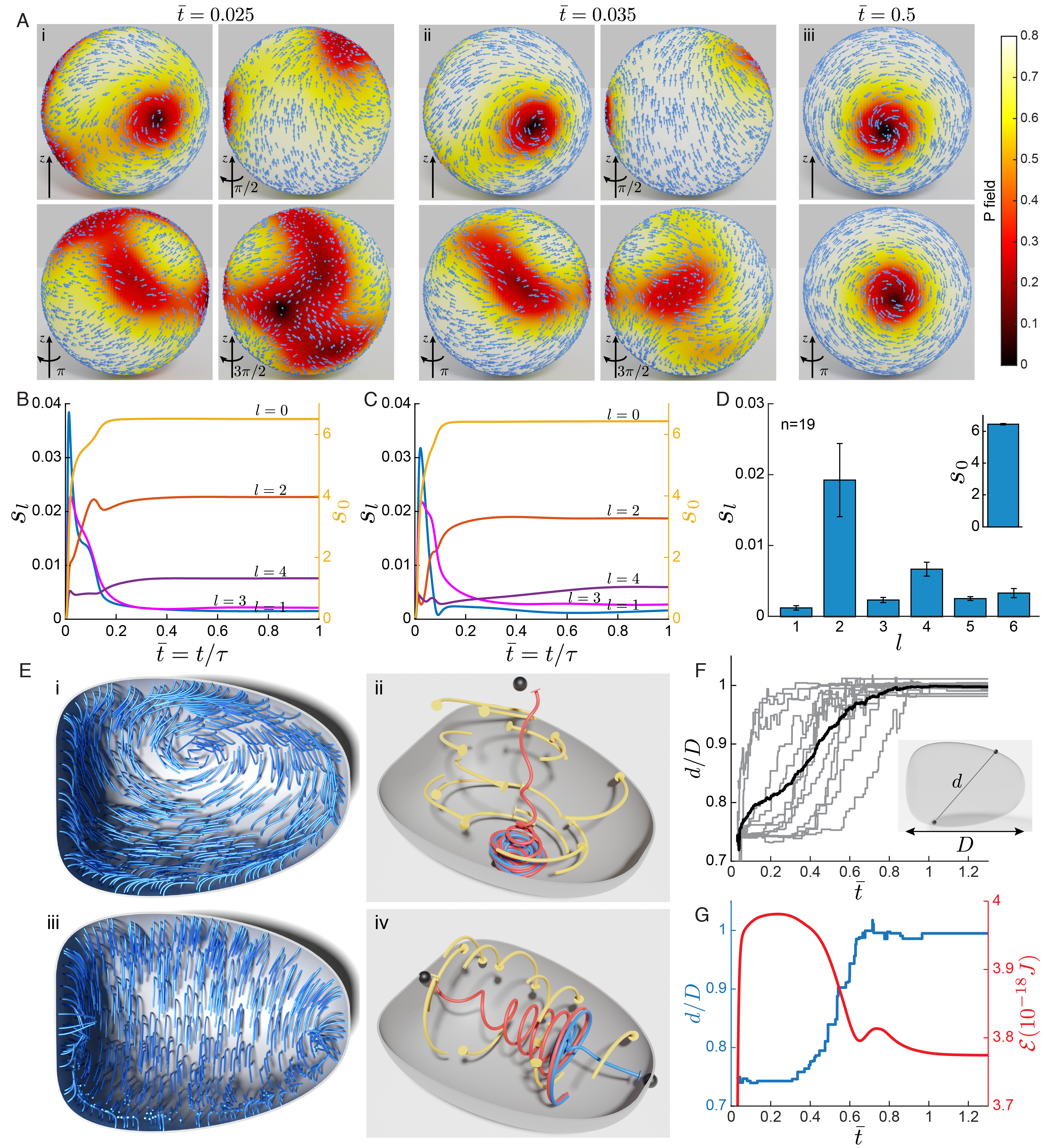}
\caption{\textbf{The structure of the streaming flow is robust.}
(A)~Microtubule surface polarity vectors $\mathbf{p}_i$ (shown as arrows) and the scalar polar order parameter $P$ from multiple views. Views (i) and (ii) are at early time ($t=0.025 \tau_r$, $t=0.035 \tau_r$) and (iii) at long time ($t=0.5 \tau_r$). $P=1$ (bright) represents highest level of local alignment and $P=0$ (dark) represent lack of local alignment. (B-C)~Angular power $s_{l}$ of  spherical harmonic coefficients of the $P$ field, $s_l(t)=\sum_{m=-l}^{m=l}|\tilde{P}_{lm}(t)|^2/(2l+1)$, for two simulations with microtubules positioned at the same place but with different initial configuration. The right and left axes represent $l=0$ and $l \neq 0$, respectively. (D)~Bar graphs representing the mean steady-state value of $s_l$ ($l=0$ shown in inset) for $n=19$ such simulations. Error bars show standard deviation. Similar statistics are found by sampling different microtubule anchoring distributions (supplementary Fig. S5). (E)~Microtubule configurations from a simulation with cell geometry similar to \textit{Drosophila} oocyte from an early~(i) and a late~(iii) time point, using control parameters $\bar{\sigma}$ and $\bar{\rho}$ of case II. Integrated streamlines of the instantaneous flow field at the corresponding times (ii, iv). The disks represent the point of origin of the streamline. The streamlines in red and blue has point of origin near the regions with the lowest microtubule polarity $P$, denoted by black spheres. (F) Normalized distance between the two defect centers as a function of time for 13 simulations in the oocyte geometry, as indicated in the inset (gray, individual simulations, black, average). (G) Distance between the defect centers (blue) and total elastic energy $\mathcal{E}$ of microtubules (red) as a function of time for simulation in (E).
 }
\label{FIG4}
\end{figure}

\smallskip
{\it Robust emergence of twisters:} We studied the emergent states over different initial conditions and for different realizations of statistically uniform placement of microtubules. This showed self-organized streaming to be very robust, with the main variation being the orientation of the twister in the cell.

The evolution towards this state can be followed through the dynamics of a surface polar order-parameter, $P(\mathbf{y})$ ($\mathbf{y}$ on the surface), of microtubules, obtained by averaging together surface polarity vectors over a (sliding) surface disk $\nu$ centered on $\mathbf{y}$: $P=|\mathbf{P}|$, where $\mathbf{P}(\mathbf{y},t)=\langle\mathbf{p}_i(t)\rangle_\nu$.
Low polar order is achieved by cancellation of anti-aligned $\mathbf{p}_i$ vectors, or by their originating microtubules being close to orthogonal to the surface (both of which are evinced near defect centers; see Figs.~\ref{FIG2}E,F). For initially straight microtubule beds, $P(\mathbf{y},t=0)\equiv 0$. This unstable state quickly evolves into a state with multiple spatially complex regions with high polar order (Fig.~\ref{FIG4}A, supplementary Movies 6,7). Each of these high order regions is contributing to the cytoplasmic flow through which these regions compete and interact. The low-order regions gradually sharpen, as high order regions expand and merge, finally resolving into the axisymmetric swirling state whose axis joins the two opposing defect centers of low polar order.

While the orientation of this axis depends on the fine-grained details of microtubule placement and initial conditions, the final flows and order parameter fields all evolve towards the same basic attractor. Their strikingly similar progression of self-organization the streaming twister state can be readily appreciated by examining the $P$-field's angular power spectra: $s_l(t)=\sum_{m=-l}^{m=l}|\tilde{P}_{lm}(t)|^2/(2l+1)$ (components of the spherical harmonic power spectrum, with $l$ the polar mode index). As the system evolves towards the twister state, both even and odd $l$ modes initially grow and then (i) either saturate for even $l$ while being dominated by the global $l=0$ mode, or (ii) relax back to relatively small amplitudes for odd $l$ modes. The relaxation timescale of these modes is similar to the collective relaxation timescale $\tau_c$. We consistently observed similar dynamics and end-states for simulations, where anchoring points or microtubule initial conditions were varied~(Fig.~\ref{FIG4}B-D, supplementary Fig. S5). 



{\it Model dynamics in an oocyte geometry:} Like cows, oocytes actually have a variety of shapes, being very roughly distended ellipsoids with approximate symmetry around their anterior-polar axis. How might cell nonsphericity affect twister formation and dynamics? To investigate this, we simulated motor-loaded microtubule beds anchored within an axisymmetric "oocyte-shaped" cell (supplementary information). The dynamics is first familiar and then surprising. Beginning from straight microtubules and using the case II values for $\bar{\sigma}$ and $\bar{\rho}$, the progressive growth and coarsening of domains with bent but aligned microtubules is again observed. This process leads again to a twister [Fig.~\ref{FIG4}E(i) and supplementary Movie 8] that sits askew the cell, respecting no obvious geometric symmetry, with the structure of its interior flows a geometric perturbation of the rotational plus bitoroidal flows found in spheres [Fig.~\ref{FIG4}E(ii)]. Evolving from different samples of microtubule anchoring points leads to twisters at differing orientations, usually tilted within the cell (Fig.~\ref{FIG4}F). However, we find that in all cases for this cell shape, the newly formed twisters slowly reorient into alignment with the anterior-posterior cell axis [Figs.~\ref{FIG4}E(iii),F] while preserving the basic interior flow structure [Fig.~\ref{FIG4}E(iv)]. 

What drives this reorientation? In this system, energy is stored in the elastic deformations of the microtubule bed, measured by its total elastic energy
\begin{equation}
    \mathcal{E}(t)=\sum_{i=1}^{N}\frac{E}{2}\int_0^{L^i} ds [\kappa^i(s,t)]^2,
\end{equation}
where $\kappa^i$ is microtubule curvature. The elastic energy is driven by motors performing work on the system, and dissipated by viscous and drag forces. While the energy is initially zero, as microtubules are initially straight, $\mathcal{E}$ begins a rapid rise as microtubules collectively bend, and reaches its maximum as the twister forms (Fig.~\ref{FIG4}G). Rather than lingering there, as it would if the system were in steady-state, $\mathcal{E}$ immediately begins decreasing and the twister axis starts its rotation. Moving away from the first twister state and into alignment with the anterior-posterior axis, $\mathcal{E}$ generally decreases and finally relaxes into a steady-state of reduced elastic energy. 

\smallskip
{\it Discussion:} Taken together, our results underscore the robustness of cytoplasmic streaming that emerges from hydrodynamic interactions among cortically anchored microtubules loaded with cargo-carrying motors. Fine-tuning is not required; as long as microtubule density and motor activity are within the wide domain of parameters that corresponds to stable streaming, self-organization takes care of the rest, establishing a cell-spanning twister. Naming such vortices twisters begs comparison with the more familiar kind. Tornadoes are inertia dominated, and have highly localized vortical cores, maintained by axial swirl, away from which flow velocities decay. Our zero Reynolds number twisters have velocities rising from the center, ala solid body rotation, and reflect a precise balance between active surface driving and viscous dissipation. 

Our model makes several interesting predictions. As discussed, our twister states are volumetric structures which predicts that standard microscopy imaging should show a variety of flows, depending on how the imaging volume intersects the flow structure. This is consistent with our own experimental observations. Further, our simulations are of statistically uniform microtubule beds in axisymmetric cell shapes. Thus, our twisters have no preferred direction of rotation with respect to axes of cell symmetry. Our experimental flow reconstructions likewise showed no evidence for rotational biases, clockwise or counter-clockwise, with respect to the anterior-posterior cell axis (supplementary Figs.~S3C,D). Our simulations show small secondary streaming flows originating from defect centers. While the rotational flow seems very robust these secondary flows may not be, and may be affected by various cellular inhomogeneities; As yet, our microscopy imaging volumes are currently insufficient to make a clear comparison. Very interestingly, simulated twisters in ooctyte shapes show a slow reorientation -- on the long $\tau_r$ timescale -- towards axisymmetrization. While this is a challenging prediction, requiring long-time, stable imaging of the oocyte, we are pursuing new observations. The nature of this dynamics also suggests the existence of a manifold of twister states, towards which the system is rapidly attracted, and upon which twisters slowly move towards the axisymmetric state, which we believe to be the state of minimum energy (supplementary Fig.~S6A). This picture is consistent with simulations in spheres, where elastic energy also shows a rapid peaking then decay (supplementary Fig.~S6B). But, the apparent overshoot is now far smaller, with subsequent dissipation towards a state of barely different energy (supplementary Fig. S6A), suggesting a slow twister dynamics driven by statistical details of microtubule placement, rather than cell shape. The nature of transition to swirling and to axisymmetry requires further exploration, but it is interesting to note that changes in viscosity in our model do not lead to state bifurcations but only changes the time-scale upon which dynamics occurs.

What function might a twister serve? Before the onset of streaming, diffusion and motor-driven transport are the main means by which different classes of RNAs are transported and anchored to the the anterior, dorsal, and posterior regions of the oocyte. Importantly, several gene products guard the oocyte against precocious streaming, since it would interfere with the localization of transcripts needed for embryonic patterning ~\cite{Quinlan2016,manseau1989}. Once these factors are stably localized, the oocyte switches to a streaming strategy for intracellular transport. The appearance of streaming may reflect its role in the uptake of yolk, the main source of protein in the embryo. Yolk proteins arrive to the future egg cell via internalization, after binding to a specific receptor which localizes to the oocyte plasma membrane shortly before streaming onset~\cite{schonbaum2000}. The streaming flow might be used to efficiently disperse of the arriving yolk throughout the ooplasm. This idea is consistent with the fact that yolk proteins are known cargoes of kinesin-1 motors walking on cortically anchored microtubules~\cite{Lu2016}, and with our preliminary analyses of the mixing capabilities of self-organized intracellular twisters.

\smallskip
{\it Acknowledgments:}
We thank Brato Chakraborti, Jasmin Imran Alsous, Elizabeth Gavis, and Raymond Goldstein for extensive and useful discussions, and Arvin Farhadifar for generously sharing his Blender expertise. We acknowledge support by NIH grants R01GM134204 (SYS), R35GM131752 (VIG), and NSF grant DMR-2004469 (MJS). Stocks obtained from the Bloomington Drosophila Stock Center, supported by NIH grant P40OD018537, were used in this study. 

\smallskip
{\it Author Contributions:}
MJS, SYS, and VIG designed the research. SD, RF, GK, and RB contributed to simulation software development and simulation data analysis. WL and ML designed and performed the experiments. SD, RF, and MJS developed image processing software and analysis of experimental data. SD, RF, SYS, and MJS prepared the manuscript. All authors contributed to its editing.

\smallskip
{\it Data Availability:}
Simulational and experimental data sets generated during the current study are available from the corresponding author upon reasonable request. 

\smallskip
{\it Code Availability:}
A publicly available,
and elaborated, version of the {\it SkellySim} codebase, used to generate the simulations, is available at \\ https://github.com/flatironinstitute/SkellySim.

\newpage 
\section*{Materials and Methods}
\renewcommand{\theequation}{S\arabic{equation}}
\subsection*{Simulation of microtubules in closed geometries}

We briefly outline our method of simulation, the technical details of which appear in \cite{Nazockdast2017}, which combine slender-body theory and boundary integral methods for solving the Stokes equations. A publicly available, and elaborated, version of the underlying code, {\it SkellySim}, is available at \cite{SkellySim}.

In broad strokes, we simultaneously solve the coupled equations of motion, Eqs. (1) and (2) in the main text, for the fluid and the immersed microtubules confined in the cellular volume $\Omega$ and clamped at the boundary $\Gamma$. Due to the linearity of the Stokes equations, we can write the fluid velocity at $\mathbf{x}$ in $\Omega$ as  $\mathbf{u}(\mathbf{x})=\mathbf{u}^{mt}(\mathbf{x})+\mathbf{u}^{\Gamma}(\mathbf{x})$, with $\mathbf{u}^{mt}(\mathbf{x})=\sum_i \mathbf{u}^{i}(\mathbf{x})$ the superposition of velocities induced by forces and conformations of each microtubule $i$, and $\mathbf{u}^{\Gamma}(\mathbf{x})$ the consequent backflow velocity induced by the no-slip condition at the confining boundary $\Gamma$.
The velocities $\mathbf{u}^{mt}$ and $\mathbf{u}^{\Gamma}$ are expressed in terms of two fundamental solutions to the Stokes equations, the Stokeslet tensor $\mathbf{G}$ (a second-rank tensor), and the Stresslet $\mathcal{T}$ (a third-rank tensor):
\begin{equation}
\label{eqs1}
\mathbf{G}(\mathbf{x})
   =\frac{1}{8\pi\mu}
   \frac{\mathbf{I}+\hat{\mathbf{x}}\hat{\mathbf{x}}}
   {|\mathbf{x}|};~~~
   \mathcal{T}(\mathbf{x})
   =\frac{-3}{4\pi\mu}
   \frac{\hat{\mathbf{x}}\hat{\mathbf{x}}\hat{\mathbf{x}}}
   {|\mathbf{x}|^2}.
\end{equation}
$\hat{\mathbf{x}} = \mathbf{x}/\|\mathbf{x}\|$. Slender-body theory for the Stokes equations gives that, to leading (logarithmic) order in the slenderness ratio $\epsilon$, the velocity induced by a microtubule is given as a line integral of the distribution of the Stokeslets along its centerline: 
\begin{equation}
	\mathbf{u}^i(\mathbf{x}) = \int_0^{L^i} \mathbf{G}\left(\mathbf{x}-\mathbf{r}^i(s')\right)\mathbf{f}^i(s')\, ds'.
\end{equation}
where $\mathbf{f}_i$ is the internal elastic force that a microtubule exerts upon the fluid (see main text).

The second contribution, $\mathbf{u}^\Gamma$, accounts for the no-slip condition taken upon $\Gamma$, and is expressed as a surface convolution of the Stresslet over $\Gamma$ with an unknown density $\mathbf{q}$:
\begin{equation}
\label{u_b}
\mathbf{u}^{\Gamma}(\mathbf{x}) = 
\int_{\Gamma} dS_y \, \mathbf{n}(\mathbf{y}) \cdot \mathcal{T}(\mathbf{r})\cdot\mathbf{q}(\mathbf{x}),
\end{equation}
where $\mathbf{r}=\mathbf{x}-\mathbf{y}$, and $\mathbf{n}$ is the outward normal vector to $\Gamma$. In the parlance of integral equations, this is a double-layer representation. In such a representation, taking the limit $\mathbf{x}\rightarrow\Gamma$ of Eq.~(\ref{eqs1}) and applying the no-slip condition, $\mathbf{u}=0$, generates a well-conditioned Fredholm integral equation of the second kind for $\mathbf{q}$:
\begin{equation}
\label{Fredholm}
-\frac{1}{2}\mathbf{q}(\mathbf{x}) + 
\int_{\Gamma}\, dS_y \mathbf{n}(\mathbf{y}) \cdot \mathcal{T}(\mathbf{r})\cdot\mathbf{q}(\mathbf{y}) 
+ \int_{\Gamma}\, dS_y [\mathbf{n}(\mathbf{x})\mathbf{n}(\mathbf{y})]\cdot \mathbf{q}(\mathbf{y}) = -\mathbf{u}^{mt}(\mathbf{x}),~~~~
\mathbf{x}\in\Gamma.
\end{equation}
Here, the last term of the RHS is added to complete the rank (i.e. make it uniquely invertible) of the integral equation. This term does not change the velocity $\mathbf{u}^\Gamma$ but does fix a constant in the pressure field \cite{Nazockdast2017}.
In Eq.~(1) of the main text the background velocity for microtubule $i$ is given by $\bar{\mathbf{u}}^i(\mathbf{x})=\sum_{j\neq i}\mathbf{u}^j(\mathbf{x})$, i.e., the flows induced by all other microtubules. At each time, the unknown field to determine for microtubule $i$ is its tension field $T^i$ which enforces inextensibility. This condition generates, through Eq.~(1) of the main text, integro-differential equations for all the $T^i$s. Solution of the coupled system for $(\mathbf{q},T)$ allows calculation of the microtubule velocities $\mathbf{X}_t^i(s,t)$. 

The integrodifferential operators along the centerlines of the microtubules are discretized using $4^{th}$-order finite differences. For the time stepping, we use an adaptive explicit/implicit backward time-stepping scheme, which maintains accuracy while removing high-order stability stiffness constraints from the bending term. This results in a dense linear system of equations which we solve using GMRES with block-diagonal preconditioners. We accelerate computing the hydrodynamic interactions using the Fast Multipole Method~\cite{Greengard1987}. The complexity per time-step scales with the total number of discretization points, on microtubules and the cell surface.

\subsection*{Biophysical and numerical parameters of simulations}
{\it Biophysical:} In our simulations, we chose the length of all microtubules to be $L=20~\mu$m -- on the longer side if growing from dynamical instability -- and having bending rigidity $E=20~\mbox{pN}~\mu\mbox{m}^2$ \cite{gittes1993}. Given a microtubule diameter of $\sim 20~$nm gives $\epsilon\sim 5\cdot 10^{-4}$. If the cell is spherical it is of radius $R=100~\mu$m, taken in this abstracted shape as the typical size for stage 10 \textit{Drosophila} oocytes. For an "oocyte"-shaped cell, whose construction is described below, the length is $150~\mu$m, and width $108~\mu$m. The immersing fluid is taken as Newtonian with viscosity $\mu=1$Pa~s \cite{Ganguly2012}. 

The relaxation time of a microtubule is estimated as $\tau_r=\eta L^4/E\sim16,000~$s. For comparison, this is somewhat less than the duration of stage 10 of Drosophila development -- approximately 10~hr$=$36,000~s -- where large scale streaming flows first appear. We note that streaming persists into stage 12. Generally we have $(\tau_c,\tau_m)=(\tau_r/\bar\rho,\tau_r/\bar\sigma)$. For beating case I, this gives $(\tau_c,\tau_m)\approx(3200,180)$s, while for the streaming case II, we have $(\tau_c,\tau_m)\approx(1066,355)$s.

{\it Numerical:} Microtubules are clamped orthogonally to the inner surface of a model cell. Microtubules are placed randomly on the cellular surface with a uniform probability, with their placements filtered to ensure that any two microtubules are not closer than distance $\Delta=0.1L$ from each other. Each microtubule is discretized with 64 points. The maximum allowable time-step is $\Delta t=0.16~$s, much smaller than any of faster time-scales $\tau_c$ or $\tau_m$.

\subsection*{Classification of microtubule dynamics in simulations}

We established the phase diagram of the model based on the dynamics and shape of the microtubules in long-term simulations. In the stable phase, microtubules remain unperturbed and normal to the surface; in the beating phase, microtubules' shapes continuously change with time; in the streaming phase, microtubules attain steady deformed shapes. To classify the simulations into these three phases, we first measured the normalized positional variance of the microtubule's free end
 \begin{equation}
\delta^2_i=\frac{\left\langle(\mathbf{X}^i(L)-\langle\mathbf{X}^i(L)\rangle)^2\right\rangle}{L^2},
 \end{equation}
where the averaging is over a period of $\Delta t=10^{-3}\tau_r$. If $\delta^2_i<0.1$, we consider the microtubule shape time-independent; otherwise, its shape is dynamic. For a simulation, if more than 90$\%$ of microtubules are dynamic, we classify it as the beating phase; otherwise, it belongs to stable or streaming phases. To distinguish between these two phases, we measured the projection of the microtubule end-to-end vector as
\begin{equation}
    a_i=1-\hat{\mathbf n} \cdot \frac{\mathbf X_i(L)-\mathbf X_i(0)}{|\mathbf X_i(L)-\mathbf X_i(0)|}
\end{equation}
If $a_i<0.05$, the microtubule is considered normal to the surface, otherwise, it is deformed. For a simulation, if more than 90$\%$ of microtubules are normal to the surface, we classify it as the stable phase. Otherwise, we classify it as the streaming phase. 

\subsection*{Analytical approximation of the streaming flow in sphere}
We approximated the flow in simulations of the sphere as a superposition of a swirling flow, $\mathbf{u}_{\rm s}$, and an axisymmetric bitoroidal flow, $\mathbf{u}_{\rm t}$~\cite{stone1991}
\begin{dmath}
\begin{aligned}
 &\mathbf{u}_{\rm ana}=\mathbf{u}_{\rm s}+ \mathbf{u}_{\rm t}\\
 &\mathbf{u}_{\rm s}(r,\theta,\phi)= \Omega \frac{r}{R}\sin (\theta)  \hat {\boldsymbol \phi}\\
&\mathbf{u}_{\rm t}(r,\theta,\phi)=\frac{A}{R^3} \left[  r(r^2- W^2)(1-3\cos^2 \theta) \hat{\boldsymbol r}+r(5 r^2- W^2)\cos(\theta)\sin (\theta) \hat  {\boldsymbol \theta}\right],
\end{aligned}
\end{dmath}
where, $r$, $\theta$, $\phi$ are the radial, polar, and azimuthal coordinates in the sphere, and $\hat{\boldsymbol{r}}$, $\hat{\boldsymbol{\theta}}$, and $\hat{\boldsymbol{\phi}}$ are the unit vectors in the respective directions. The three parameters of the model are $\Omega$, the strength of the swirling flow, $A$, the strength of the bitoroidal flow, and $W$ is the radius associated with the bitoroidal flow. We fit our simulations of spherical geometry to this flow by minimizing $\displaystyle \xi=\int (\mathbf{u}_{\rm ana}-\mathbf{u}_{\rm sim})^2 dV$, where the integration is over the sphere volume. For the minimization, we use the gradient descent algorithm for six free parameters, including three angles, to align the axis of the flow in simulation to the $z$ axis. We found that $\Omega=(100.2 \pm 3.0)$ nm/s, and $A=(5.5\pm 1.2)$ nm/s. The ratio of the strength of the swirling flow to the toroidal one is $\Omega/A\sim 20$.

\subsection*{Simulation in oocyte-shaped geometry}

To study the model in a geometry similar to \textit{Drosophila} oocyte, we construct a surface of revolution as
\begin{eqnarray}
\label{oocyte_coords}
 &&X=Dx \quad \nonumber\\
 &&Y=Dr \cos (\phi) \quad \nonumber\\
 &&Z=Dr \sin (\phi),
\end{eqnarray}
where $x \in ( 0, 1 )$, $\phi \in [ -\pi, \pi )$, and $r=\frac{{T}x^{p_1}(1-x)^{p_2}}{2^{(1-p_1-p_2)}}$(see ~\cite{Stoddard2017}). $l$ is the oocyte length, $T$ sets the aspect ratio of the oocyte, and the parameters $p_1 \in [0, 1]$ and $p_2 \in [0, 1]$ determine the local curvature of the oocyte. In our simulations, we chose $D=150  \mu m$, $T=0.72$, $p_1=0.4$, and $p_2=0.2$. 

\subsection*{Live imaging of the \textit{Drosophila} oocyte}
Young mated female adults were fed with dry active yeast for 16-18 hours and dissected in Halocarbon oil 700 (Sigma-Aldrich, Cat: H8898) as previously described~\cite{Lu2018, Lu2016}. Samples were imaged within 1 hour after dissection, using Nikon W1 spinning disk confocal microscope (Yokogawa CSU with pinhole size 50 µm) with Photometrics Prime 95B sCMOS Camera or Hamamatsu ORCA-Fusion Digital CMOS Camera, and a 40X 1.25 N.A. silicone oil lens, controlled by Nikon Elements software. 3D time-lapses were acquired every 10 seconds at $1\mu m$/step.

Flies were maintained on standard cornmeal food (Nutri-Fly Bloomington Formulation, Genesee, Cat: 66-121) supplemented with dry active yeast (Red Star) at room temperature (24 – 25°C). The following fly stocks were used in this study: mat $\alpha$tub-Gal4[V37] (III, Bloomington \textit{Drosophila} Stock Center:7063)~\cite{Lu2021}; UASp-F-Tractin-tdTomato (II, Bloomington stock center:58989)~\cite{Spracklen2014, Lu2021}; GFP::$\alpha\rm{tub}$~\cite{grieder2000}.

\subsection*{Reconstruction of 3D velocity field from live imaging}
Here, we describe the steps for 3D reconstruction of the velocity field and measurement of microtubule orientation from experimental images. First, we reconstructed the 3D oocyte periphery, then measured the 2D cytoplasmic velocity field for each z-plane using particle image velocimetry, and then used it to reconstruct the 3D velocity field. We also measured the local orientation of microtubules using a linear filter for texture analysis.


\subsubsection*{3D Reconstruction of the oocyte periphery}
We developed an active contour method~\cite{kass1988, farhadifar2014} to partially reconstruct the 3D geometry of the oocyte from volumetric images of F-actin. We first segmented the oocyte periphery for each z-plane and then used these to reconstruct the 3D oocyte surface. In short, for the middle z plane, we provided a closed curve, $\Tilde{\Gamma}(s)$, which serves as the initial guess for the active contour method. The shape of the oocyte in this z-plane is given by minimizing the cost function
\begin{equation}
    E[\Gamma(s)]=\oint_{\Gamma(s)}\frac{\alpha}{2}|\Gamma_{ss}|^2ds+\iint_R(I_N(x,y)-\beta)dxdy 
\end{equation}
where $\Gamma_{ss}=\partial^2\Gamma/\partial s^2$. The first term accounts for the smoothness of the contour, and the second term accounts for the interaction of the contour with the image, where $\beta$ is set such that the contour expands if it is far from the periphery. The image intensity, $I_N(x,y)$, is the normalized smoothed gradient of the F-actin image. We then used the segmented shape of the oocyte in this z-plane as an initial guess to segment the oocyte periphery in consecutive z-planes.

\subsubsection*{2D particle image velocimetry}
Particle image velocimetry (PIV) is a common technique for inferring the local velocity of the fluid by measuring the displacement of tracer particles between two consecutive time points. In brightfield microscopy images of \textit{Drosophila} oocyte, lipid granule particles have high contrast relative to the cytoplasm and can serve as tracer particles to measure local cytoplasmic velocity. We developed a platform to perform PIV on brightfield microscopy images of \textit{Drosophila} oocytes. A key piece of our software is using the contrast-limited adaptive histogram equalization method to enhance the contrast of the brightfield images~\cite{Zuiderveld1994}. To accurately measure the velocity in the complex geometry of the oocyte, we combined fast Fourier transform-based PIV (FFT-based) on a square grid within the interior of the oocyte and correlation-based PIV near the periphery. 

For FFT-based PIV, square boxes of 100 pixels with 20 pixels spacing far from the oocyte periphery were taken~\cite{Willert1991, Thielicke2014}. For each box, we calculated the Fourier transform of the image intensity for two constitutive time points, $\tilde{I}_t(u,v)$ and $\tilde{I}_{t+1}(u,v)$, calculated the Hadamard product of one with the complex conjugate of the other, $\tilde{I}_H=\tilde{I}_t\circ\tilde{I}_{t+1}^*$, and set the displacement in that box as the position of the maximum of the inverse Fourier transform of $I_H(\Delta x, \Delta y)$ ~(Fig. \ref{FIGS2}a blue; pixel size, $0.260\mu m$).

For the correlation-based PIV for points near the periphery, we first constructed grids with shapes derived from the oocyte outline as follows: we chose $N$ evenly spaced points on the periphery, and for each point, we constructed $M$ evenly spaced points on a line connecting the center of mass of the oocyte cross-section to that point. By connecting each set of points, we constructed $N\times M$ grids~(Fig. \ref{FIGS2}a). For each grid, we calculated the displacement by finding the maximum of the correlation function 
\begin{equation}
	H(\Delta_x,\Delta_y)= \int \int_A I_t(x,y)I_{t+1}(x+\Delta_x,y+\Delta_y) {\rm d}x  {\rm d}y, 
\end{equation}
where $I_{t}$, and $I_{t+1}$, are the mean subtracted intensity, and the integration is over the area of the grid. If there are multiple local maxima, we choose the one giving a smooth displacement field between the neighboring grids. Finally, we use interpolation to estimate the planar components of the velocity field, $u_x(x,y,z)$ and $u_y(x,y,z)$, on a regular grid across the oocyte using these two displacement fields (Fig.~\ref{FIGS2}a).

\subsubsection*{Approximation of the out-of-plane velocity}

We measured the out-of-plane component of the velocity field, $u_z(x,y,z)$, by assuming the incompressibility of the cytoplasm and the impermeability of the oocyte boundary ($\Gamma$) (on the timescales of microscopy) and solving
\begin{equation}	
	\nabla\cdot\mathbf{u}=0;\quad \mathbf{u}\cdot\hat{\mathbf{n}}|_{\Gamma}=0 
\end{equation}
where $\hat{\mathbf{n}}$ is the surface normal vector. To do so, we numerically solve the ODE
\begin{equation}
    \frac{\partial u_z}{\partial z}=
    -\left(\frac{\partial u_x}{\partial x}
    +\frac{\partial u_y}{\partial y} \right) ,
\end{equation}
with the boundary condition $u_z=-(n_x u_x+n_y u_y)/n_z$ at the oocyte periphery~(Fig. \ref{FIGS2}b right).

\subsection*{Estimation of microtubule orientation field from microscopic images}
To measure the local microtubule orientation, we use a Gabor filter, which is a linear filter for texture analysis~\cite{Jain1991}. It allows examination of any specific frequency content in the image in a given direction. The inputs of the filter are its wavelength and orientation, and the outputs are the magnitude and phase response to the filter. We use 3-pixel wide wavelength for angles $\theta \in (0,\pi]$ with interval $\pi/180$, and for each angle, we calculate the magnitude response in grids of $20\times 20$ pixels. We set the grid orientation as the angle with the largest magnitude response and grid magnitude as the value of the magnitude response to that angle~(Fig. \ref{FIGS4}). 

\subsection*{Estimation of the positions of the defect}
As illustrated in Fig. 2E in the main text, near the defect centers microtubules are relatively straight and normal to the surface. To find the defect positions in the oocyte simulation, we sort  all the microtubules in the descending order of $a_i$ (equation (S6)), the length of projection of the end-end vector on the surface. We assign the first defect center to be at the location where the microtubule with lowest $a_i$ is clamped. From the rest of the microtubules, we find the one with next lowest $a_i$, which is at least 40 $\mu m$ away from the first defect and assign it's clamping position to be the second defect.

\section*{Supplementary figures}
\setcounter{figure}{0}
\makeatletter 
\renewcommand{\thefigure}{S\@arabic\c@figure}

\begin{figure}[tbh!]
\centering
\includegraphics[width=18cm]{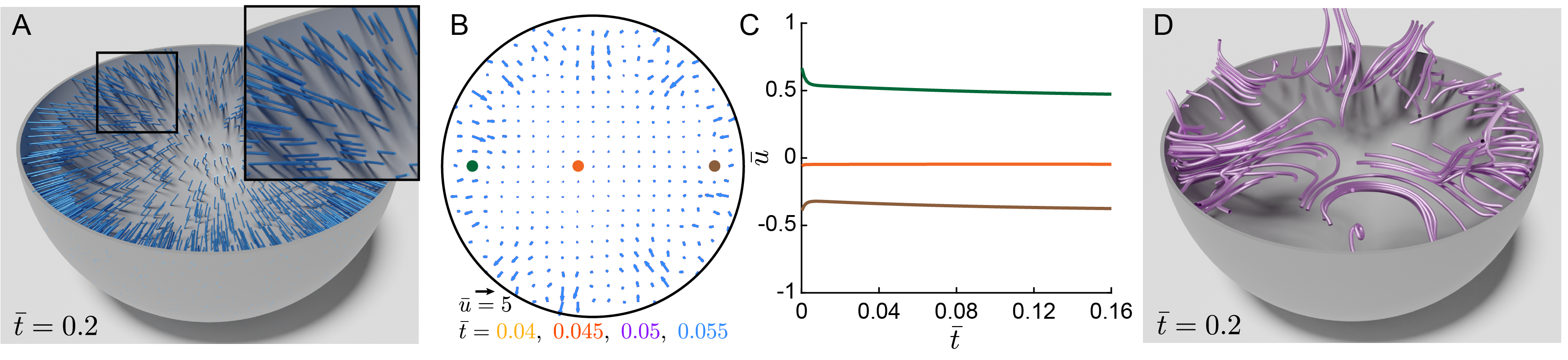}
\caption{Simulation in the stable regime, using $\bar{\rho}=15$ and $\bar{\sigma}=5$. (A) Cut-away view of instantaneous microtubule configurations in the spherical cell. Inset shows very slightly bent microtubules. (B) 2D projection of velocity field in the sectioning equatorial plane of (A), at four time points. (C) the azimuthal velocity component at the three points labelled in the equatorial plane, as a function of $\bar t$. Note the very small magnitude of the velocities. (D) 3D streamlines integrated from the (low magnitude) 3D velocity field for a simulation for a stable regime with parameters .}
\label{FIGS1}
\end{figure}

\begin{figure}[tbh!]
\centering
\includegraphics[width=16cm]{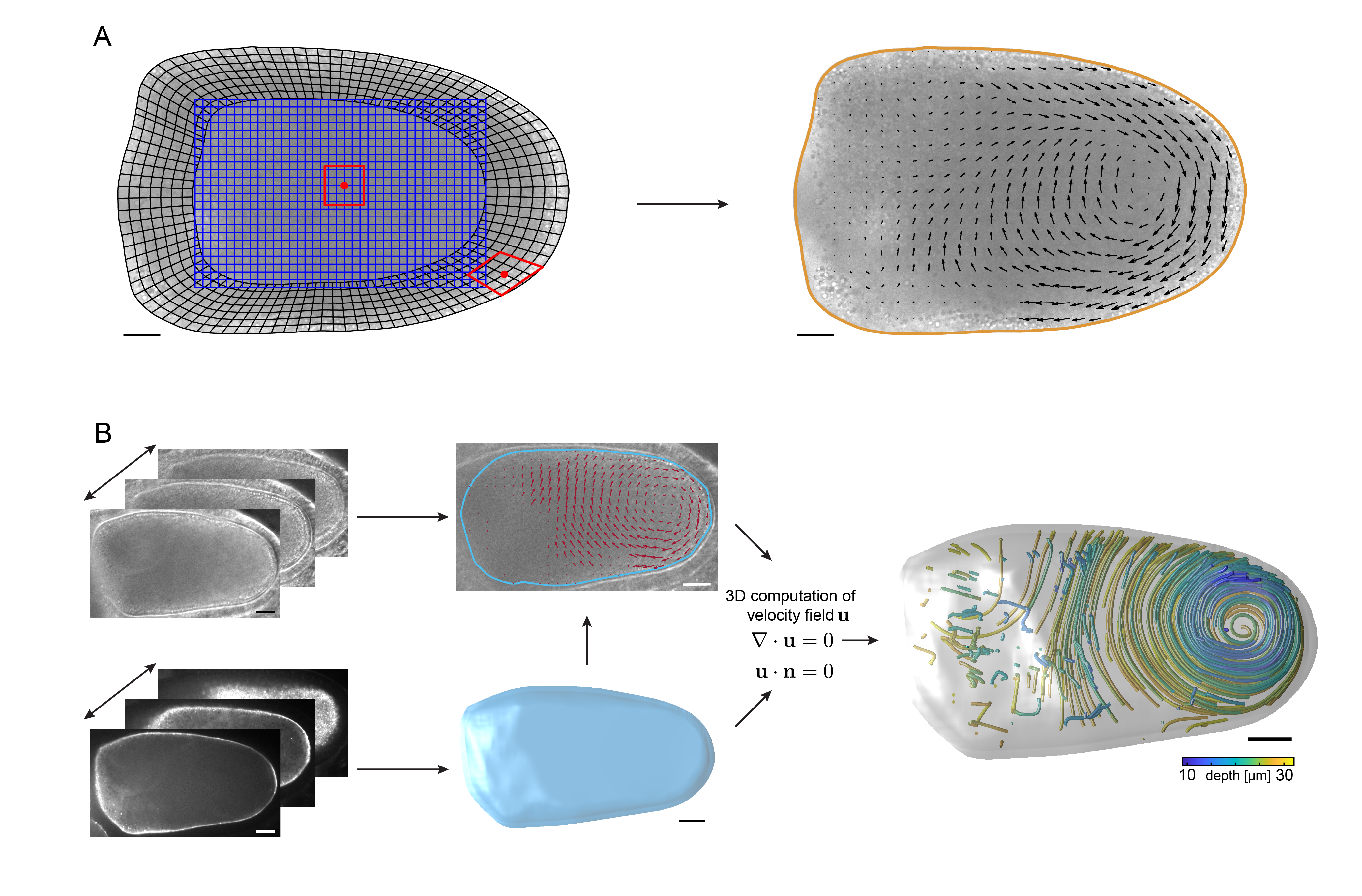}
\caption{Elements of the 3D flow measurements in the \textit{Drosophila} oocyte. (A) Left panel shows the overset grid generated by combining a Cartesian mesh for points far from oocyte boundary (blue) and a boundary-conforming mesh constructed from segmentation of the oocyte periphery (black). Right panel shows the 2D cytoplasmic velocity field in the corresponding slice. Scale bar, $25\mu m$. (B) Steps for the reconstruction of 3D flow structure from live images: 3D segmentation of oocyte periphery, 2D PIV on individual slices, computation of out-of-plane velocity component by assuming flow incompressibility, and 3D flow streamline reconstruction. Images are as seen in the microscope, with anterior at the left and posterior at the right. Scale bar, $25\mu m$.}
\label{FIGS2}
\end{figure}

\begin{figure}[tbh!]
\centering
\includegraphics[width=18cm]{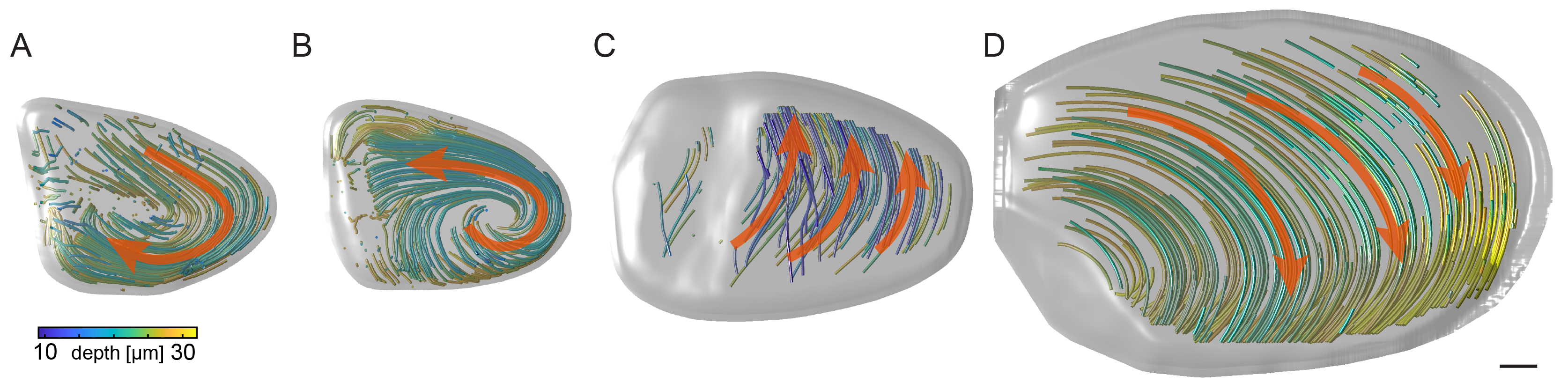}
\caption{Examples of flow structures observed within different experimental oocytes. (A,B) show vortical "twister-like" flows, while (C,D) show likewise commonly observed cross-cell streaming. Red arrows show flow direction. All images are as seen through the microscope, and are oriented with anterior at the left, and posterior at the right. Note that if interpreted through the lens of a confined twister within the cell, (C) could be interpreted as counter-clockwise motion around the oriented anterior-posterior axis, while (D) takes the opposite direction. Scale bar, $25\mu m$.}
\label{FIGS3}
\end{figure}

\begin{figure}[tbh!]
\centering
\includegraphics[width=8.5cm]{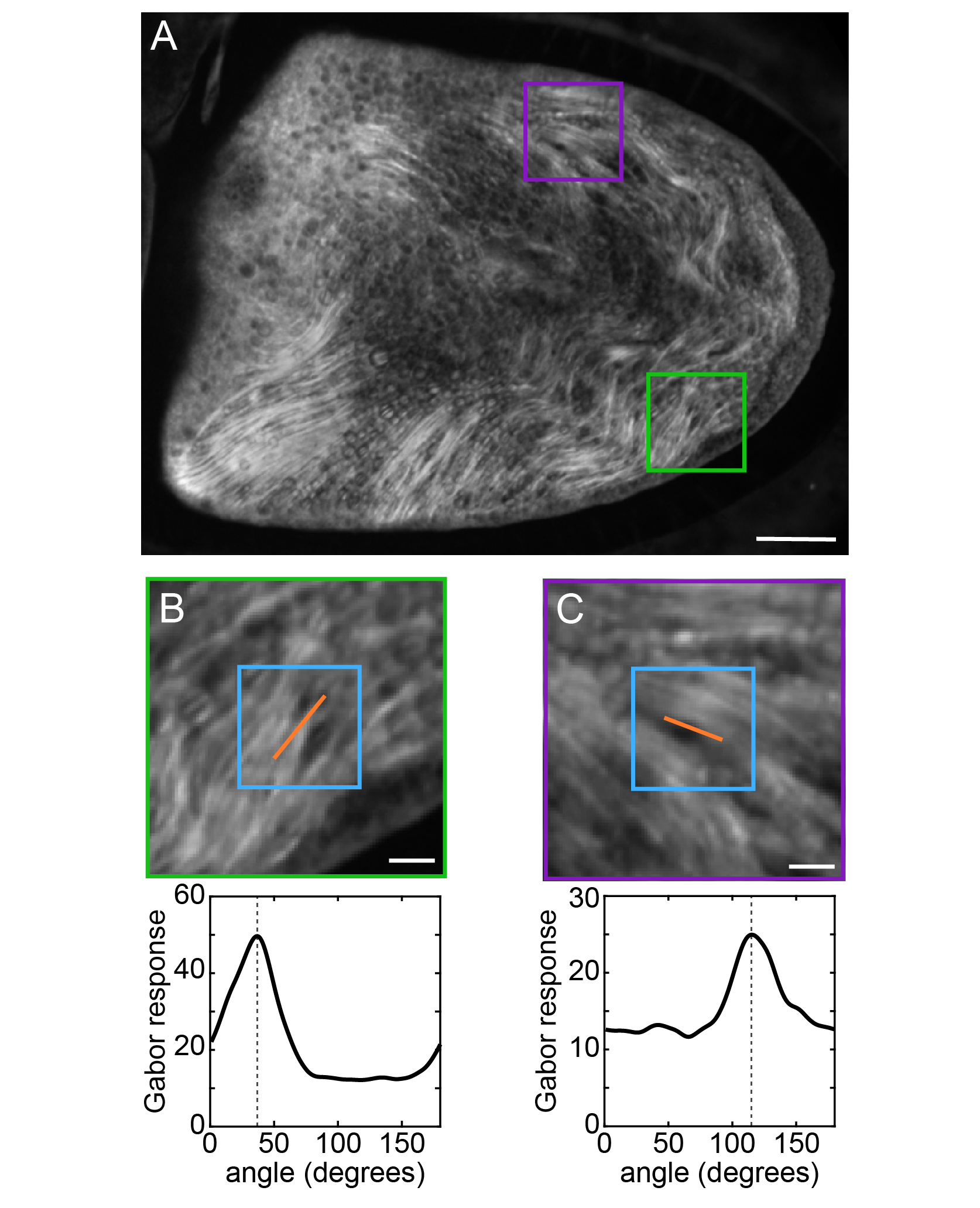}
\caption{Measurement of microtubule orientation from live images of \textit{Drosophila} oocytes. (A) A 2D projection of microtubules tagged with maternally derived GFP-$\alpha$tub. Scale bar, $25 \mu m$. (B)-(C) B corresponds to the green box in A, while C corresponds to the purple. Orange lines: microtubule orientation of peak power response of the Gabor filter applied to a $20\times20$ pixel box (blue). The graphs are the Gabor response as a function of angle relative to the x-axis. Images are as seen in the microscope, with anterior at the left and posterior at the right. Scale bar, $5 \mu m$.}
\label{FIGS4}
\end{figure}

\begin{figure}[tbh!]
\centering
\includegraphics[width=14cm]{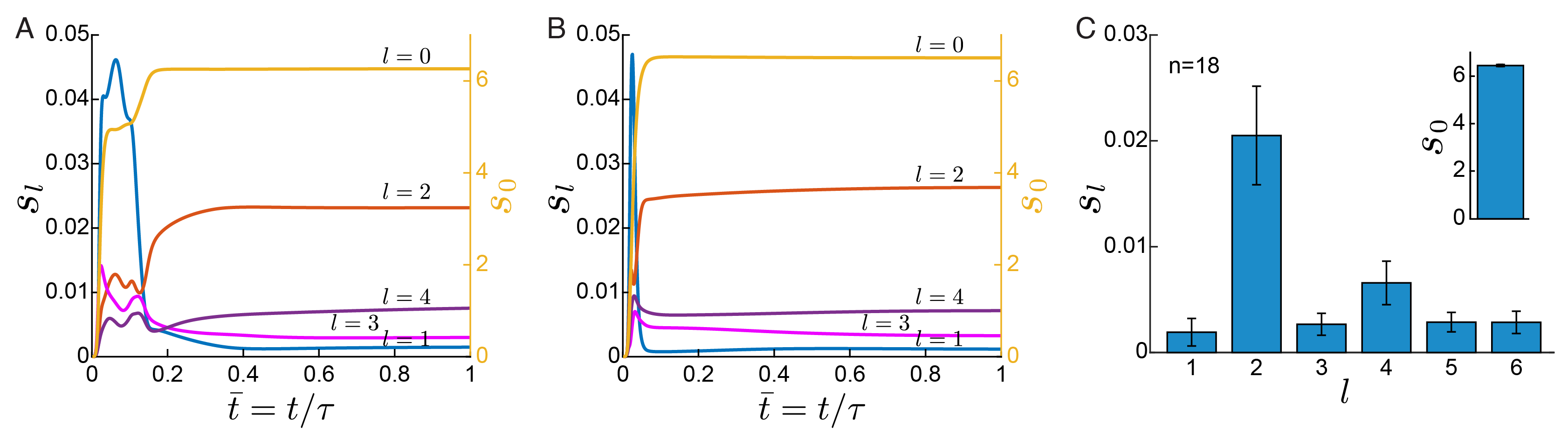}
\caption{Robustness of twister formation to microtubule distribution. (A-B)~Angular power $s_{l}$ of  spherical harmonic coefficients of the $P$ field, $s_l(t)=\sum_{m=-l}^{m=l}|\tilde{P}_{lm}(t)|^2/(2l+1)$, for two simulations with different microtubule anchoring distributions but with same (straight) initial configuration. The right and left axes represent $l=0$ and $l \neq 0$, respectively. (C)~Bar graphs representing the mean steady state value of $s_l$ ($l=0$ in insets) for 18 such simulations (error bars show standdard deviations).}
\label{FIGS5}
\end{figure}

\begin{figure}[tbh!]
\centering
\includegraphics[width=14cm]{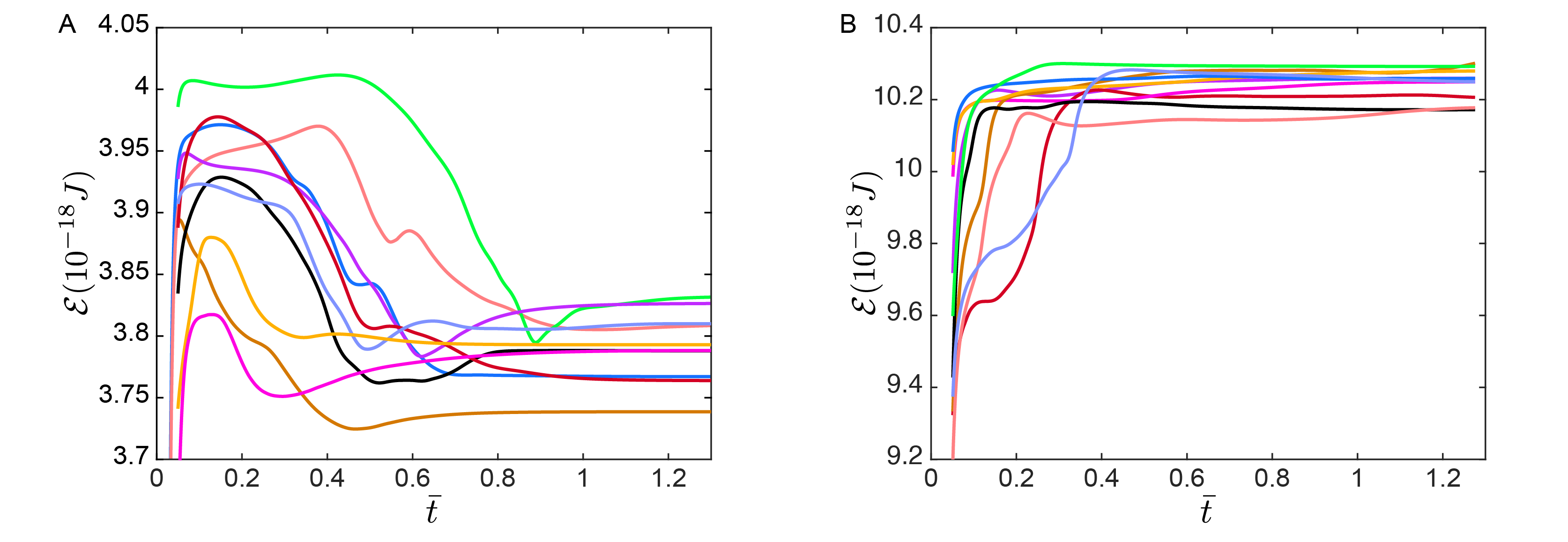}
\caption{Evolution of the elastic energy in spherical and oocyte-shaped cells. (A) total elastic energy $\mathcal{E}$ for 10 independent realizations (different microtubule placements) of case II in a spherical cell geometry as a function of adimensional time $\bar{t}=t/\tau_r$. Note that all simulations evolve towards twisters of similar elastic energy. (B) 10 such simulations, again using case II parameters, in the oocyte geometry of Fig.~4E of the main text. All show an overshoot in the elastic energy, followed by relaxation to a lower energy axisymmetric state. These simulations correspond to those shown in Fig.~4F of the main text demonstrating reorientation to axisymmetry.}
\label{FIGS6}
\end{figure}

\end{document}